\documentclass[floats,floatfix,amssymb,prd,twocolumn,superscriptaddress,nofootinbib,preprintnumbers]{revtex4-1}

\usepackage{subcaption,ragged2e}
\DeclareCaptionJustification{justified}{\justifying}
\usepackage{amssymb,amsmath,verbatim,mathtools,needspace,enumitem,etoolbox,graphicx,microtype,afterpage,bm}
\usepackage{tikz}
\usetikzlibrary{shapes.misc, positioning, arrows.meta}
\usepackage{framed}
\usepackage{soul}
\usepackage{hyperref}
\hypersetup{colorlinks, citecolor=bluscuro, linkcolor=black, urlcolor=bluscuro}
\definecolor{rossos}{cmyk}{0,1,1,0.55}
\definecolor{bluscuro}{rgb}{0.15, 0.2, .85}
\definecolor{bluchiaro}{cmyk}{1,.3,0.,0.1}
\definecolor{ForestGreen}{rgb}{0.13, 0.55, 0.13}
\definecolor{TLGreen}{RGB}{50, 164, 49}
\definecolor{TLOrange}{RGB}{231,180,22}
\definecolor{TLRed}{RGB}{204,50,50}

\usepackage{multirow,pifont,lmodern,float,tabularx,booktabs}
\usepackage[normalem]{ulem}
\usepackage[all]{hypcap}
\usepackage[T1]{fontenc}
\usepackage[utf8]{inputenc}
\usepackage[capitalise]{cleveref}
\allowdisplaybreaks

\newcommand{\Np}{N_{\rm p}}
\newcommand{\Ns}{N_{\rm s}}
\newcommand{\Npatch}{N_{\rm patch}}

\newcommand{\be}{\begin{equation}}
\newcommand{\ee}{\end{equation}}
\renewcommand{\d}{{\rm d}}

\newcommand{\unipd}{Dipartimento di Fisica e Astronomia ``G. Galilei'', Università degli Studi di Padova, via Marzolo 8, I-35131 Padova, Italy}
\newcommand{\infnpd}{INFN, Sezione di Padova, via Marzolo 8, I-35131 Padova, Italy}

\newcommand{\iem}{
Instituto de Estructura de la Materia (IEM), CSIC, Serrano 121, 28006 Madrid, Spain}

\begin{document}

\title{
Are PTA measurements sensitive to gravitational wave non-Gaussianities?
}

\author{Chiara Cecchini}
\email{chiara.cecchini@csic.es}
\affiliation{Department of Physics, University of Trento, via Sommarive 14, 38122 Povo (TN), Italy}
\affiliation{Trento Institute for Fundamental Physics and Applications-INFN, via Sommarive 14, 38122 Povo (TN), Italy}
\affiliation{\iem} 
\author{Jonas El Gammal}
\email{jonas.el.gammal@kbfi.ee}
\affiliation{Como Lake Center for Astrophysics, Department of Science and High Technology, University of Insubria, via Valleggio 11, I-22100, Como, Italy} 

\author{Gabriele Franciolini}
\email{gabriele.franciolini@unipd.it}
\affiliation{\unipd}
\affiliation{\infnpd}

\author{Mauro Pieroni}
\email{mauro.pieroni@csic.es}
\affiliation{\iem}

\begin{abstract}
Observing non-Gaussianities in the timing residuals of Pulsar Timing Arrays (PTAs) has recently attracted attention as a potential discriminator between astrophysical and cosmological origins of the observed Gravitational Wave (GW) signal. In this work, we show that even in an idealized signal-dominated setup, after decorrelating the data to avoid spurious detections, statistical tests applied to PTA data cannot distinguish between a Gaussian and a non-Gaussian amplitude distribution of the GWB in a model-agnostic way. In particular, without making strong assumptions on the GW spectrum or the properties of the population, the sensitivity to any distinctive non-Gaussian feature is washed out.
\end{abstract}

\maketitle
\preprint{CERN-TH-2025-241} 
\hypersetup{linkcolor=bluscuro}

\textbf{\emph{Introduction --}}
Pulsar Timing Arrays (PTAs) track the pulses of millisecond pulsars, whose rotational stability enables accurate predictions of their times-of-arrival at Earth. Gravitational Waves (GWs) induce correlated timing residuals; for an isotropic GW Background (GWB), the correlation pattern averaged over realizations follows the Hellings--Downs (HD) curve~\cite{Hellings:1983fr}, a robust fingerprint of a GWB.

Several PTA collaborations reported evidence for a signal consistent with nanoHertz (nHz) GWs, exhibiting correlations compatible with the HD prediction~\cite{NANOGrav:2023gor,EPTA:2023fyk,Reardon:2023gzh,Xu:2023wog,Miles:2024seg}, whose origin remains unknown: the commonly invoked explanation relies on a population of Supermassive Black Hole Binaries (SMBHBs)~\cite{Rajagopal:1994zj,Jaffe:2002rt}, but various early-Universe mechanisms remain viable given current uncertainties~\cite{Madge:2023cak,NANOGrav:2023hvm,EPTA:2023xxk,Figueroa:2023zhu,Ellis:2023oxs,Caprini:2024lxj,Moore:2021ibq}.

Distinguishing astrophysical from cosmological sources using spectral information is challenging, as both classes of models yield broad and overlapping predictions. This has motivated the search for complementary observables. A promising direction is to search for GWB anisotropies: cosmological mechanisms predict a nearly isotropic GWB, whereas SMBHBs are expected to produce $\mathcal{O}(1\text{--}20\%)$ anisotropies, $C_{\ell}/C_0$, in the lowest multipoles~\cite{Mingarelli:2013dsa,Mingarelli:2017fbe,Sah:2024oyg}. Their detection would strongly favor an astrophysical origin and therefore has become a central target for PTAs~\cite{Taylor:2013esa,NANOGrav:2023tcn,Ali-Haimoud:2020iyz,Hotinli:2019tpc,
Gardiner:2023zzr,Lemke:2024cdu,Konstandin:2024fyo,Depta:2024ykq,Domcke:2025esw,Gersbach:2025mhj}.

Another approach is probing the statistical properties of the GWB amplitude. For SMBHBs, the signal arises from the superposition of a discrete and finite set of quasi-monochromatic binaries, leading to deviations from Gaussianity when few bright sources dominate (e.g.,~\cite{Becsy:2023qul}). In contrast, cosmological GWBs in the nHz band are generated no later than when the Universe had a temperature $T \sim 10^2\,\mathrm{MeV}$, around the QCD epoch~\cite{Schwarz:1997gv,Caprini:2018mtu,Franciolini:2023wjm}. The number of disconnected Hubble patches contributing to the signal is $\Npatch \sim 10^{18}$~\cite{Allen:1996vm}. The overall signal results from the superposition of many incoherent contributions, and the central limit theorem (CLT) drives the strain toward strong Gaussianity~\cite{Bartolo:2018rku}.

While several works investigated non-Gaussianities (NGs) in PTA data~\cite{Xue:2024qtx,Bernardo:2024uiq,Falxa:2025qxr,Lamb:2025niq,Kuntz:2026usl,Ciprini:2026pvz,Raidal:2026ezm}, we systematically study PTA's ability to measure them. To isolate the signal statistics, we adopt an idealized, signal-dominated setup, neglecting measurement noise and timing-model systematics. 

Timing residuals are generally correlated because they integrate signals from overlapping sky regions through the broad detector response function.
We account for this by performing a linear transformation (rotation and rescaling in pulsar-pulsar space) of the timing residuals to the basis identified by the principal components of the HD correlation matrix. 
In this basis---under the null assumption that the signal is Gaussian and isotropic---the variables are uncorrelated and have unit variance.

We show that any NG test assuming isotropy and estimating the power (i.e., variance) directly from the data loses sensitivity and cannot reject the null hypothesis. In doing so, we account for the full distribution of the rotated timing residuals (i.e., we do not restrict to only the lowest-order moments).
We interpret this effect as the manifestation of the CLT: as each pulsar's residuals receive contributions from many sky directions modulated by the response function, their distribution rapidly converges towards a Gaussian.

Our analysis is based on the Python code \texttt{fastPTA},~\footnote{\href{https://github.com/Mauropieroni/fastPTA}{github.com/Mauropieroni/fastPTA}} extended with \texttt{fastropop},~\footnote{\href{https://github.com/jonaselgammal/fastropop}{github.com/jonaselgammal/fastropop}} a package for the generation of SMBHB populations.

\textbf{\emph{PTA response to GWs --}}
PTAs measure the timing residual $\delta t_I$ induced by GWs on the arrival time of a signal emitted by a pulsar $I$, at distance $D_I$ in direction $\hat p_I$~\cite{Romano:2016dpx,Burke-Spolaor:2018bvk,Taylor:2021yjx}. Expanding the metric perturbation $h_{ab}$ in plane waves of frequency $f$, polarization $P=\{+,\times\}$, and sky direction $\hat\Omega$, one obtains
\begin{align}
\delta t_I &=
\sum_P
\int_{{S^2}} {\rm d}^2\hat\Omega
\int_{-\infty}^{\infty} {\rm d}f\,
\tilde h_P(f,\hat\Omega)\,
R_I^P(f,\hat\Omega)\,
\frac{e^{2\pi i f t}}{i\,2\pi f},
\label{eq:tim_res_integrated}
\end{align}
where $\tilde h_P(f,\hat\Omega)$ are the Fourier coefficients and we have introduced the response function 
\begin{align}
R_I^{P}(f,\hat\Omega)
&\equiv
\frac{\hat p_I^a \hat p_I^b\, e^P_{ab}(\hat\Omega)}
     {2\left(1+\hat p_I\!\cdot\! \hat\Omega\right)}
\left[1-e^{-i 2\pi f D_I\, (1+\hat\Omega\cdot\hat p_I)}\right],
\label{eq:response}
\end{align}
and $e^P_{ab}(\hat\Omega)$ are the polarization tensors.
For GWBs, the timing residuals have vanishing expectation value\footnote{Notice that this is an expectation value over statistical GWB realizations, but not over pulsar distributions.}, $\langle \delta t_I \rangle = 0$. Thus, the information is encoded in the cross-correlations, with expectation values
\begin{align}\label{eq:covariance}
\widetilde{C}_{IJ} (f) &\equiv \langle \widetilde{\delta t}_I (f) \widetilde{\delta t}_J^* (f) \rangle = \frac{1}{(2\pi)^2 f^2  } \, R_{IJ}(f)  {\cal P} (f),
\end{align}
where $\widetilde{\delta t}_I (f)$ is the Fourier transform\footnote{Since the observation time is finite, we can only measure a discrete set of frequency bins, which are affected by spectral leakage and exhibit non-negligible correlations across nearby bins. For simplicity, we neglect these effects, the inclusion of which can only strengthen our conclusion.} of $ \delta t_I$, and we have substituted the expectation value for the cross-correlations for a stationary, isotropic, and unpolarized GWB 
\begin{equation}
\langle \tilde h_P(f,\hat\Omega) 
\tilde h_{P'}^*(f',\hat \Omega') \rangle
= \frac{{\cal P} (f)}{4} 
\delta_{PP'}
\delta^2(\hat \Omega,\hat \Omega')
\delta(f-f'), 
\label{eq:hab_corr}
\end{equation}
where ${\cal P}(f)$ is the GW power spectral density (PSD), and we have defined 
\begin{align}\label{eq:Gamma_IJ}
R_{IJ}(f)
\equiv
\sum_P
\int \frac{{\rm d}^2\hat\Omega}{4\pi}\,
R_I^P(f,\hat\Omega)\,
R_J^{P*}(f,\hat\Omega).
\end{align}
Since the exponential term in the brackets of \cref{eq:response}, called ``pulsar term'', is rapidly oscillating, it results in 
\begin{equation}
    \left[1-e^{-i 2\pi f D_I\, (1+\hat\Omega\cdot\hat p_I)}\right] \left[1-e^{i 2\pi f D_J\, (1+\hat\Omega\cdot\hat p_J)}\right] \simeq (1 + \delta_{IJ})\,,
\end{equation}
which factorizes out. Then, \cref{eq:Gamma_IJ} reduces to the well-known HD correlation~\cite{Hellings:1983fr}, which we denote as $\Gamma_{IJ}$.

\textbf{\emph{Cosmological and astrophysical signal statistics --}}
In general, for any given frequency, sky direction, and polarization, the signal can be expressed as 
\begin{equation}
\tilde h_P(f,\hat\Omega) = \sum_{n=1}^{ \Ns } A_{P,n}(f,\hat\Omega_n) e^{i\phi_{P,n}(f,\hat\Omega_n)}\delta^2(\hat\Omega, \hat \Omega_n),
\end{equation}
where $A_{P, n}(f,\hat\Omega_n)$ and $\phi_{P,n}(f,\hat\Omega_n)$ are the amplitude and phase of the $n$-th contribution located along the direction $\hat{\Omega}_n$, and $\Ns$ corresponds to the number of incoherent Hubble patches (independent GW sources) for a cosmological (astrophysical) GWB.
For $\Ns \gg 1$ with random phases uniformly distributed in $[0,2\pi)$, the CLT drives the real and imaginary parts of $\tilde h_P(f,\hat\Omega)$ toward Gaussian distributions. 

Note that~\cref{eq:Gamma_IJ}, which imposes the pulsar-pulsar covariance to be (approximately) proportional to the HD correlation, assumes an isotropic GWB. Thus, any sizable deviation from isotropy will induce a different correlation pattern among the pulsars. In the following, we take a Gaussian and isotropic signal as our null hypothesis and test for deviations from this pattern.

While most cosmological GWBs are highly Gaussian, deviations from Gaussianity can, in principle, arise in special cases (see, e.g.,~\cite{Maldacena:2011nz,Namba:2015gja,Unal:2018yaa,Cai:2018dig,Bartolo:2018qqn,Anninos:2019nib,Adshead:2021hnm,Yuan:2023ofl,Bartolo:2019oiq,Bartolo:2019yeu,Kumar:2021ffi,Li:2023qua,Li:2024zwx}). However, propagation in the perturbed Universe washes out such primordial NGs~\cite{Bartolo:2018evs,Margalit:2020sxp}, unless one focuses on ultra-squeezed configurations (e.g.~\cite{Dimastrogiovanni:2019bfl}).
Propagation effects can additionally imprint anisotropies, e.g. via Shapiro time delay~\cite{Contaldi:2016koz,Jenkins:2018nty,Bartolo:2019yeu,Cusin:2022cbb,Tasinato:2023zcg}, but are proportional to the large-scale power spectrum, which is very small. 
Finally, known anisotropic contributions, such as the one induced by motion~\cite{Smoot:1977bs} (expected to be small in any case), can be included by augmenting the covariance accordingly~\cite{Mingarelli:2013dsa,Taylor:2013esa,Gair:2014rwa,NANOGrav:2023tcn}. Thus, a cosmological GWB can be well approximated by our null hypothesis.

In contrast, an astrophysical GWB from SMBHBs can have very different properties, as the source population is discrete and finite. First, the number of binaries contributing at a given frequency $f$ follows Poisson statistics with mean $\langle N(f) \rangle \propto f^{\alpha}$, where $\alpha=-8/3$ for circular, GW-driven binaries~\cite{Phinney:2001di,Sesana:2008mz} and different values of $\alpha$ arise in more realistic models including environmental effects and eccentricity~\cite{Enoki:2006kj,Kocsis:2010xa,Chen:2016zyo,Kelley:2016gse,Huerta:2015pva,Taylor:2015kpa,Sampson:2015ada,Burke-Spolaor:2018bvk}. 
Poisson fluctuations in $N/\langle N\rangle$ directly translate into signal variance~\cite{Sesana:2008mz,Rosado:2015epa}.
Since only a single realization is observable, this statistic cannot be measured without detailed knowledge of the underlying population~\cite{Agazie:2024jbf}. 
Moreover, the signal amplitude is generically expected to deviate from a Gaussian distribution~\cite{Lamb:2024gbh}. 
\emph{In this work, the NG in the amplitude distribution is the quantity of interest}.
A second deviation from the null hypothesis comes from the fact that each SMBHB emits a deterministic, quasi-monochromatic signal with well-defined amplitude, frequency, and phase evolution from a given sky location~\cite{Sesana:2004sp,Sesana:2012ak,Burke-Spolaor:2018bvk}. However, the HD curve, our null hypothesis, assumes an \textit{isotropic} signal~\cite{Allen:2022dzg,Romano:2023zhb}. Any astrophysical GWB dominated by a few loud sources will, in principle, violate our assumption.

\textbf{\emph{Decorrelating the data --}}
From~\cref{eq:covariance}, it is clear that in a PTA, since the covariance is (approximately) proportional to the HD correlation, which is not diagonal, the data $\widetilde{\delta t}_I (f) $ are not independent draws. A Gaussianity test applied directly to them would mistake these correlations for a genuine departure from Gaussianity in the signal.

\begin{figure}
    \centering
    \includegraphics[width=0.85\linewidth]{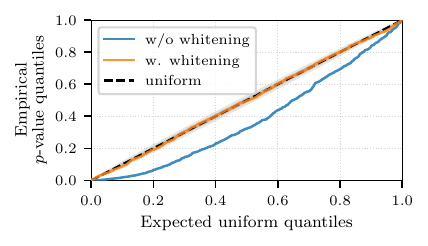}
    \caption{QQ plot of $p$-values under the isotropic Gaussian null hypothesis for diagonal powers before (blue) and after (orange) whitening, with theoretical expectation (dashed-black) and $68\%$, $95\%$ confidence intervals (gray bands).}
    \label{fig:whitening_qq}
\end{figure}

To illustrate this, we generate mock datasets with $\Np=67$ samples drawn from a multivariate normal distribution with zero mean and a known covariance matrix. We then test the Gaussian hypothesis using the Kolmogorov-Smirnov (KS) test (see Appendix~\ref{app:statistical_tests}).
The resulting Quantile-Quantile (QQ) plot of $p$-values (\cref{fig:whitening_qq}, blue line) exhibits a statistically significant deviation from the diagonal, demonstrating that correlations lead to a spurious rejection of the null hypothesis.

Therefore, the first step in testing for NG is to remove correlations using the covariance structure of the null hypothesis. This operation consists of a whitening transform (see below) that, under the null hypothesis, converts correlated samples into uncorrelated unit-variance variables. After such a whitening procedure, the data (orange line in~\cref{fig:whitening_qq}) are indeed compatible with the null hypothesis. A similar approach is standard in other cosmological analyses, e.g.\ of the Cosmic Microwave Background~\cite{Frommert:2012}.

For simplicity, we consider a single frequency bin\footnote{The PTA signal spans multiple frequency bins that are not independent due to finite observation time. The whitening procedure can be generalized to multiple bins to keep track of frequency-frequency correlations and applied to the full covariance matrix across pulsars and frequencies \cite{Depta:2024ykq,Crisostomi:2025vue}.} at $f=f_i$, setting $d_I \equiv \widetilde{\delta t}_{I}(f=f_i)$, with $I=1,\ldots,\Np$. Under the null hypothesis, we can perform an eigenvalue decomposition of the covariance (i.e., $\propto R_{IJ}$ up to frequency factors) as
\begin{equation}
\widetilde{C}_{IJ} =
\frac{1}{(2 \pi f)^2}R_{IJ} {\cal P} (f) = M_{IK} \Lambda_{KL} M^\dagger_{LJ},
\label{eq:eigendecomp}
\end{equation}
where $\Lambda_{KL} = \lambda_K \delta_{KL}$ is diagonal with positive eigenvalues $\lambda_K$, and $M$ is unitary. Defining
\begin{equation}
W_{IJ} \equiv (\Lambda^{-1/2} M^\dagger)_{IJ}, \qquad w_I \equiv W_{IK} d_K,
\label{eq:Wdef}
\end{equation}
by construction yields
\begin{equation}
\langle w_I w_J^* \rangle = (W \widetilde{C} W^\dagger)_{IJ} = \delta_{IJ}\ .
\label{eq:whitecond}
\end{equation}
Thus, under the Gaussian and isotropic null, the $w_I$ are independent complex Gaussian draws.

After whitening, deviations from the null hypothesis can arise from two effects discussed previously: {\it 1)} the signal amplitude distribution is genuinely non-Gaussian, or {\it 2)} the pulsar-pulsar correlations deviate from the HD structure, e.g.\ due to anisotropy. Here, we show that PTA measurements are insensitive to the first effect.

Notice that while $R_{IJ}$ is the correct basis to whiten into, in practice, we do not have access to it as the oscillatory contribution depends on $fD \sim \mathcal{O}(10^3)$, which is highly sensitive to pulsar distance uncertainties. These are known at a few percent level at best (e.g.~\cite{EPTA2-pulsars}), which is insufficient to resolve the oscillatory phase accurately.
Therefore, in real analyses, the whitening procedure can be carried out using $\Gamma_{IJ}$, which only approximately diagonalizes the dataset.

\begin{figure}[b]
    \centering
    \includegraphics[width=\linewidth]{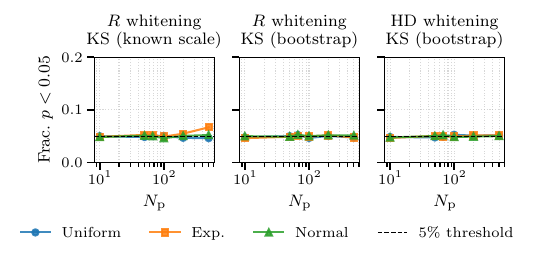}
    \caption{Rejection fractions of the KS test on isotropic injected signals drawn from different (uniform, exponential, normal) distributions and for different numbers of pulsars $\Np$. Left: Data whitened with the response function $R$ with known scale. Middle: Data whitened with $R$ with the scale estimated from the data. Right: Data whitened with the HD curve and scale estimated from the data.}
    \label{fig:isotropic_NG}
\end{figure}

\textbf{\emph{Response averaging and suppression of NGs --}}
Let us consider the case of an isotropic signal with an NG amplitude distribution in each sky patch. For this purpose, we generate artificial signal realizations (5000 per case) with the real and imaginary parts of each sky pixel drawn independently from different statistical distributions. Then, we compute the timing residuals, whiten them, apply the KS test and evaluate the fraction of realizations with $p$-value smaller than $0.05$.
 We consider two scenarios: {\it 1)} known-scale, where the variance of the Gaussian null is assumed to be known, and {\it 2)} estimated-scale, where the variance is inferred directly from the data---the realistic case for PTAs. 
 The results of this procedure, for different $\Np$ (from 10 to 500) and different amplitude distributions, are shown in \cref{fig:isotropic_NG}. It is clear that, regardless of the injection, the measured data are indistinguishable from the isotropic Gaussian null. Notice that for this setup, using $R_{IJ}$ or $\Gamma_{IJ}$ to whiten the data does not affect the results. This demonstrates that NGs in the $\tilde h_P(f,\hat\Omega)$ distribution are washed out at the timing residual level. 

Let us now remove the assumption of isotropy and consider signals that originate from the superposition of a small number of monochromatic sources with random sky positions. We choose two toy distributions for the signal amplitudes (uniform and exponential) as well as an astrophysically motivated scenario by drawing sources using the semi-analytical SMBHB model described in Appendix~\ref{app:astro_gwb}.
We generate 5000 realizations each for different combinations of source counts $\Ns$ and pulsar numbers $\Np$, compute the timing residuals, whiten them, and apply the KS test.

\begin{figure}
    \centering
    \includegraphics[width=\linewidth]{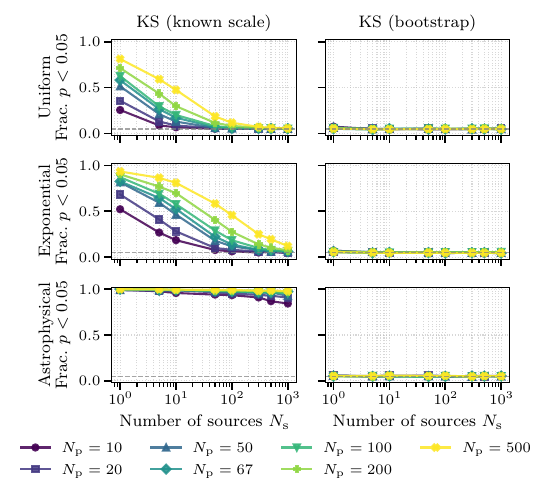}
    \caption{\label{fig:single_epoch_ks} Rejection fractions for KS tests on whitened diagonal powers. Left column: known-scale reference test; right column: bootstrap-calibrated KS test with scale estimated from the same realization. Top (middle) row: toy population with uniform (exponential) amplitudes; bottom row: semi-analytic astrophysical model with fixed source count.} 
\end{figure}

\Cref{fig:single_epoch_ks} shows the fraction of realizations whose $p$-values from the calibrated KS test are smaller than 0.05 (details in the supplemental material), assuming known and estimated scale.
Results are shown in the left and right columns of~\cref{fig:single_epoch_ks}, respectively.
The former shows that the test reacts to sparse populations as expected: sensitivity to NG increases with fewer sources (stronger NG) and more pulsars (better PTA sensitivity).
Furthermore, the heavy-tailed exponential distribution (second row) makes NGs detectable for a larger number of binaries than the uniform distribution. The consistently high rejection rate for the astrophysical population (third row) reflects its very heavy-tailed amplitude distribution~\cite{Raidal:2026ezm,Ali-Haimoud:2026sbk}, where one or very few bright binaries dominate.

The right column shows a completely different picture: in the realistic case in which the scale (i.e., the variance) of the distribution is estimated from the data, sensitivity is lost. This indicates that the $w_I$ are in fact Gaussian-distributed with some unknown scale. Thus, in the left column, the test is only assessing a scale mismatch. Once the $w_I$ are normalized by the empirical scale, no information is retained. 

\begin{figure}
    \centering
    \includegraphics[width=\linewidth]{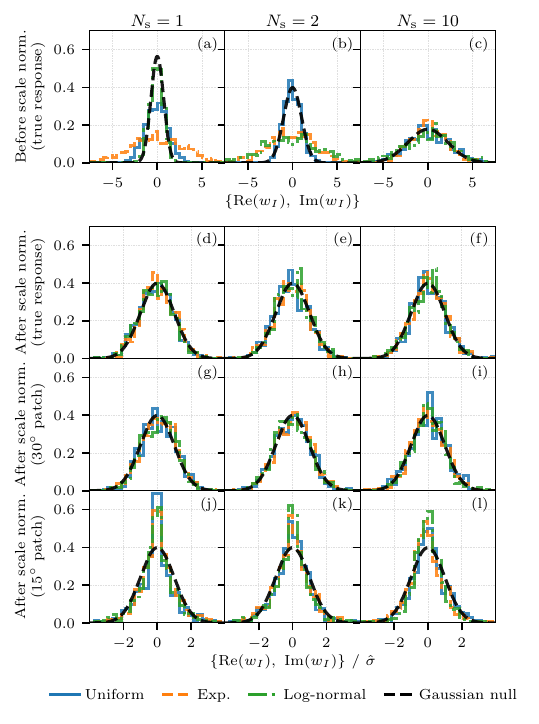}
    \caption{\label{fig:scale_estimation_cdf} Impact of scale estimation and angular resolution for $\Np=300$ pulsars and $\Ns=1,\,2,\,10$ sources (left to right). First row: amplitude PDF for uniform (solid blue), exponential (dashed orange), and log-normal (dash-dotted green) distributions, compared to the expected null (dashed black) before scale normalization, using the true PTA antenna pattern. Second row: after scale normalization. Third (fourth) row: replacing the PTA pattern with a synthetic response covering a single $30^\circ$ ($15^\circ$) sky patch (see~\cref{fig:antenna_patterns} in the supplemental material). 
    }
\end{figure}

To demonstrate this further, we focus on scenarios with few sources (the single-source case, explored further in Appendix~\ref{app:single_source}, which serves as a proxy for maximal NG). \Cref{fig:scale_estimation_cdf} shows realizations with $\Np=300$ pulsars and $\Ns=1,\,2,\,10$ (left to right), with random sky locations and amplitudes drawn from uniform (solid blue), exponential (dashed orange), and log-normal (dash-dotted green) distributions, compared to the Gaussian nulls (dashed black). Rows show the whitened distribution before~(top row) and after~(rows 2--4) scale normalization, using the realistic PTA response~(rows 1--2) and a mock single-patch response of adjustable size~(rows 3--4; see Appendix~\ref{app:antenna_patterns}).
We notice that:
\begin{enumerate}
    \item Before scale normalization (first row), the PDFs differ visibly from the null.
    \item After scale normalization (second row), the PDFs collapse to a standard Gaussian, indicating only a scale difference in the first row.
    \item For an artificially narrower antenna pattern (last two rows), the difference between the PDFs and the null is restored and becomes increasingly visible as we reduce the angular aperture of the artificial response.
\end{enumerate}

Concerning the antenna pattern, consider a few sources on the sky. For broad patterns such as the real pulsar response, the pulsars are sensitive to a sum over nearly all point sources with response-modulated amplitudes. After whitening, scale estimation erases sensitivity to the mean power on the sky. For artificially narrow response patterns, only pulsars whose beams point toward the source detect it; the rest measure no signal. This decorrelates the pulsar data, avoiding sums over multiple independent realizations of $h$, thereby avoiding CLT, and preserves information on the statistics in $w_I$ (visible as excess in the central bins of panels~j--l of~\cref{fig:scale_estimation_cdf}). A sufficiently narrow antenna pattern would thus allow detecting deviations from the null even after scale normalization---but the real pulsar response is too broad for this to yield statistical power.

Let us move on to the impact of whitening and scale normalization. Whitening is itself a linear transformation, and it decorrelates the data without affecting their statistics only if the realization-dependent correlation matrix is used. Since we only know its expectation value (and not the exact individual realization), the procedure instead produces incoherent sums over independent and undetermined variables, which, by the CLT, strongly washes out NGs. This effect is conceptually similar to, but distinct from, the averaging over multiple GW sources discussed above: here, the sum is over pulsars rather than sources. The only residual imprint is a scale mismatch that cannot be used as a discriminant without an independent GWB amplitude measurement. An artificially narrow response function suppresses both effects simultaneously: reducing the number of sources contributing to the sum and effectively decorrelating the dataset. 

In summary, while applying tests on correlated data can lead to false positives, the whitening procedure also leads to non-trivial issues. The same argument applies to \emph{any} test that probes NG, assuming isotropy and after estimating the overall scale from the same data.

\textbf{\emph{Conclusions --}}
We have studied whether tests of time-delay non-Gaussianity can distinguish a Gaussian isotropic signal from discrete-source SMBHB-like alternatives without relying on detailed prior knowledge of the SMBHB population model or a cosmological template. Our analysis is signal-dominated, neglecting noise and timing-model systematics, and should thus be seen as a best-case scenario for the constraining power of PTA experiments.

We first showed that any model-agnostic (i.e., template-free) test must be performed on whitened variables. 
The sensitivity of these variables to NG is then much
weaker than expected: for diffuse signals, the broad PTA response effectively averages over many independent signal realizations (i.e., different lines-of-sight), Gaussianizing the signal. The only remaining information is the variance of the distribution, which must be estimated from the same data and therefore cannot serve as a discriminant. For the maximally NG case of a single source, we showed analytically that scale estimation cancels the signal amplitude, leaving the data dependent only on the PTA geometry rather than the source population. 

The conclusion is that NGs are practically unmeasurable by PTAs with a model-agnostic procedure. Since our analysis assumes independent frequency bins, these conclusions do not apply to multi-frequency tests based on stronger assumptions imposing consistency across frequencies. 
We also stress that this is a statement about the measurability of NGs at the level of the timing residuals, and not about the properties of the signal itself: it does not imply that the astrophysical GWB should be modeled as Gaussian. 
Despite this result appearing rather discouraging, it opens a useful avenue: as the signal cannot produce detectable NG in PTAs, any observed deviation would necessarily point to NG effects in the timing model systematics and/or in the measurement noise. 

Finally, we stress that this does not rule out every discriminator between astrophysical and cosmological signals; it only excludes NG tests that estimate the scale from the same data. Searching for departures from isotropy remains a promising approach~\cite{Taylor:2013esa,NANOGrav:2023tcn,Ali-Haimoud:2020iyz,Hotinli:2019tpc,
Gardiner:2023zzr,Lemke:2024cdu,Konstandin:2024fyo,Depta:2024ykq,Domcke:2025esw,Gersbach:2025mhj}. Unlike NGs, anisotropies induce a distinct pulsar-pulsar correlation structure, leading to residual correlations in the data after whitening; scale normalization does not wash out this structure, so it is not subject to the limitations discussed here.

\textbf{\emph{Acknowledgments --}}
We thank Valerie Domcke, Mikel Falxa, Andrea Mitridate, and Clemente Smarra for interesting discussions and comments on the draft.
MP thanks Carlo Contaldi for the very useful discussions at an intermediate stage of this project.
We furthermore thank Bruce Allen for very useful comments on the manuscript and for the derivation of the HD-equivalent correlation function for the synthetic responses.
CC acknowledges support from the Istituto Nazionale di Fisica Nucleare (INFN) through the Commissione Scientifica Nazionale 4 (CSN4) Iniziativa Specifica ``Quantum Fields in Gravity, Cosmology and Black Holes'' (FLAG) and from Fondazione Angelo Della Riccia. 
GF~acknowledges support by the
Italian MUR Departments of Excellence grant 2023--2027
``Quantum Frontier'' and from INFN through the Theoretical Astroparticle Physics (TAsP) project.
JE acknowledges support from the Spoke 3 (INAF) of the Italian Center for SuperComputing (ICSC), funded by the European Union--NextGenerationEU program, under the grant agreement N.
C53C22000350006 (acronym Fab-HPCc). JE acknowledges the hospitality of the University of Stavanger and ETH Zürich, which provided office space during parts of this project.
The work of MP is supported by the Comunidad de Madrid under the Programa de Atracción de Talento Investigador with number 2024-T1TEC-31343. MP acknowledges the hospitality of Imperial College London, which provided office space during parts of this project. 

\bibliography{main}
\newpage

\onecolumngrid
\appendix
\section{Realistic and synthetic antenna patterns}\label{app:antenna_patterns}

\begin{figure*}[h]
    \centering
    \includegraphics[width=\textwidth]{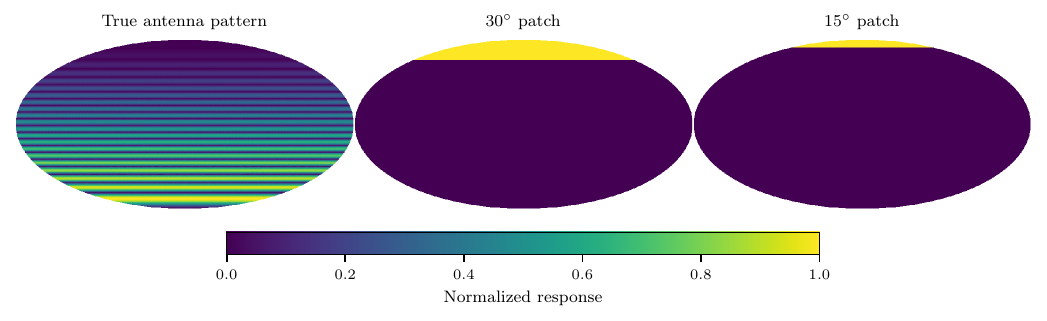}
    \caption{\label{fig:antenna_patterns} Single-pulsar sky responses for a pulsar located in the $\hat{p} = (0,0,1)$ direction shown in Mollweide projection. The left panel is the true PTA antenna pattern at $f=1 / T_{\rm obs}$ with $T_{\rm obs} = 16.03\,\mathrm{yrs}$, normalized to its maximum over the sky. The middle and right panels show the synthetic $30^\circ$ and $15^\circ$ patch responses used for the examples in~\cref{fig:scale_estimation_cdf}.}
\end{figure*}

The scale-normalized PDFs in~\cref{fig:scale_estimation_cdf} are ultimately controlled by the shape of the effective single-pulsar response on the sky. For the real PTA response, the left panel of~\cref{fig:antenna_patterns} shows the normalized single-pulsar antenna magnitude
\begin{align}
    A_I^{\rm PTA}(\hat\Omega)
    =
    \frac{
    \left[
    |R_I^+(f,\hat\Omega)|^2
    +
    |R_I^\times(f,\hat\Omega)|^2
    \right]^{1/2}
    }{
    \max_{\hat\Omega}
    \left[
    |R_I^+(f,\hat\Omega)|^2
    +
    |R_I^\times(f,\hat\Omega)|^2
    \right]^{1/2}
    },
\end{align}
with $R_I^P$ defined in~\cref{eq:response}. The synthetic responses used for comparison replace this broad PTA pattern with a top-hat patch centered on the pulsar line of sight,
\begin{align}
    A_{I,\theta_0}^{\rm patch}(\hat\Omega)
    =
    \Theta\!\left(
    \theta_0
    -
    \arccos(\hat p_I\!\cdot\!\hat\Omega)
    \right),
\end{align}
where $\Theta$ is the Heaviside step function and $\theta_0 = \{30^\circ,\ 15^\circ\}$ in~\cref{fig:antenna_patterns}. This response function corresponds to an HD-equivalent correlation curve $\Gamma^{\rm patch}(\theta_0,\cos\gamma)$ that reads: 
\begin{equation}
  \Gamma^{\rm patch}(\theta_0,\cos\gamma)
= \frac{ 2\pi
- 4 \cos\theta_0 \arccos\!\left(
\frac{\cos\theta_0 (1-\cos\gamma)}
{\sin\theta_0 \sin\gamma}
\right)
- 2 \arccos\!\left(
\frac{\cos\gamma - \cos^2\theta_0}
{\sin^2\theta_0}
\right) }{2\pi(1-\cos\theta_0) } \; ,
\end{equation}
where $\cos\gamma$ is the opening angle between two pulsars, i.e., $\cos\gamma_{IJ} \equiv \hat{p}_I \cdot \hat{p}_J$, and we have normalized to 1 for $\gamma = 0$. In the toy model, the corresponding complex patch responses are obtained by multiplying $A_{I,\theta_0}^{\rm patch}$ by independent random phases for the two polarizations before the response covariance is constructed and whitened. The true response is broad and highly overlapping, whereas the synthetic patches are progressively more localized. This is precisely why the normalized PDFs for the patch models retain more visible deviations from the $\chi^2(2)$ null: narrower responses preserve more source-location information after whitening, while the real PTA response averages that information away.

\section{Astrophysical GWB}\label{app:astro_gwb}

We model the astrophysical GWB as a superposition of distant SMBHBs. Because the binaries are far from the Earth-pulsar system compared with their GW wavelength, each source can be treated as a plane wave. For the two polarizations $h^{+}$ and $h^{\times}$ we write (see, e.g.,~\cite{Robson:2018ifk})
\begin{align}
\label{eq:hplus}
\tilde{h}_+(f) &= h(f) \frac{(1 + \cos^2 \iota)}{2} e^{i\Psi(f)} \\
\label{eq:hcross}
\tilde{h}_\times(f) &= i h(f) \cos \iota \, e^{i\Psi(f)}
\end{align}
where the amplitude $h(f)$ depends on the SMBHB parameters and the distance to the source. We have also introduced the phase $\Psi$ and the inclination angle $\iota$, i.e., the angle between the orbital angular momentum vector of the binary system and the line of sight $\hat\Omega$. When $\iota = 0$, the binary is viewed face-on. Finally, we account for the polarization angle $\psi$, which determines the orientation of the GW polarization basis with respect to a reference frame, via the random rotation
\begin{align}
\label{eq:hpolplus}
\tilde h'_+ &= \tilde h_+ \cos 2\psi - \tilde h_\times \sin 2\psi,
\\
\label{eq:hpolcross}
\tilde h'_\times &= \tilde h_\times \cos 2\psi + \tilde h_+ \sin 2 \psi. 
\end{align}

To generate the astrophysical signal, we produce an SMBHB population using the agnostic model prescription established in~\cite{Sesana:2008mz} and outlined in~\cite{EPTA:2023xxk}.
The comoving number density of merging binaries can be written as 
\be
\label{eq:d2n}
    \dfrac{\d^2n}{\d z\d \log_{10}\mathcal{M}} = \dot{n}_0 \left[\left(\dfrac{\mathcal{M}}{10^7 M_{\odot}}\right)^{-\alpha_{\mathcal{M}}}e^{-\mathcal{M}/\mathcal{M}_*}\right] \left[(1+z)^{\beta_z}e^{-z/z_0}\right]\dfrac{\d t_r}{\d z}, 
\ee
where $t_r$ is the time in the source frame. 
The five model parameters are $\theta = \{\dot{n}_0, \alpha_{\mathcal{M}}, \mathcal{M}_*, \beta_z, z_0\}$, where $\dot{n}_0$ is the merger rate per unit rest-frame time, comoving volume, and logarithmic $\mathcal{M}$ interval, and the parameter pairs $\{\alpha_{\mathcal{M}}, \mathcal{M}_*\}$ and $\{\beta_z, z_0\}$ control the shape of the $\mathcal{M}$ and $z$ distributions, respectively. 

For the analysis presented in this work, we fix the fiducial parameters to $\alpha_\mathcal{M}=0, \mathcal{M}_*=1.8\cdot 10^8 M_\odot, \beta_z=2, z_0=1.8$ and we vary the rate $\dot{n}_0$ to control the expected number of sources. The integration limits are $10^6 \leq \mathcal{M}/M_{\odot} \leq 10^{11}$ and $0\leq z \leq5$.

For simplicity, we assume that the binary evolution is dominated by the energy lost through GW emission, which means $\d f / \d t \propto f^{11/3}$. We leave generalizations to more complex models, including environmental effects, for future work. One can then compute the ensemble average of the characteristic strain as
\be
\label{eq:hcd2N}
    h^2_c(f) = \dfrac{4 G^{5/3}}{3\pi^{1/3}c^2}f^{-4/3}\int \d \mathcal{M} \int \d z (1+z)^{-1/3}\mathcal{M}^{5/3}\dfrac{\d^2 n}{\d z\d\mathcal{M}}. 
\ee
Notice that we write the equation above in terms of the observed frequency in the detector frame, which is related to the source-frame frequency as $f_r = f(1+z)$. Hence the expected scaling $h^2_c(f)\sim f^{-4/3}$.

To obtain the correct amplitude, we must compute the number of sources. Its expectation value reads
\be
\langle N \rangle = \int_{\ln f_{\mathrm{min}}}^{\ln f_{\mathrm{max}}}\d \ln f\int_{z_{\mathrm{min}}}^{z_{\mathrm{max}}} \d z\int_{\mathcal{M}_{\mathrm{min}}}^{\mathcal{M}_{\mathrm{max}}}  \d\mathcal{M}\dfrac{\d^3 N}{\d z \d \mathcal{M}\d \ln f}, 
\ee
Here, we express $\d^2n/\d z\d\mathcal{M}$ in terms of $\d^3N/(\d z \d \mathcal{M} \d \ln f_r)$ using \cref{eq:d2n} and the following definitions~\cite{Phinney:2001di}
\be
\begin{aligned}
    \dfrac{\d \ln f_r}{\d t_r} & = \dfrac{96}{5}\pi^{8/3} \dfrac{\left(G \mathcal{M}\right)^{5/3}}{c^5}f_r^{8/3},\\
    \dfrac{\d t_r}{\d z} & = \dfrac{1}{H_0(1+z)E(z)},\\
     \dfrac{\d z}{\d V_c} & = \dfrac{H_0}{4\pi c} \dfrac{E(z)}{d_L^2}(1+z)^2,
\end{aligned}
\ee
where $d_{L}$ is the luminosity distance to the source and $E(z) \equiv (\Omega_M(1+z)^3 + \Omega_k(1+z)^2+\Omega_\Lambda)^{1/2}$.
It follows that $\langle N \rangle 
\propto 
{\d N}/{\d \ln f} \propto f\,{\d t}/{\d f_r} \propto f^{-8/3} 
$.
A practical way to compute the \textit{ensemble-averaged} signal spectrum is to draw $\Ns$ samples from the distribution $\d^3N/(\d z \d \mathcal{M} \d \ln f_r)$ and add up the contributions of the individual binaries to obtain~\cite{Middleton:2020asl, Sesana:2008mz}
\be
    h_c^2(f_i) = \dfrac{\sum_k \bar{h}^2_k f_k}{\Delta f_i}, 
\ee
where the sum runs over all the $k$ sources emitting in the $i$-th frequency bin and $\bar{h}_k$ is the sky-and-inclination-averaged strain amplitude,~\footnote{$\bar{h}(f)$ is related to the GW amplitude defined in \cref{eq:hplus,eq:hcross} as $\bar{h}(f) = \sqrt{2/5} \, h(f)$.} defined as 
\be
\label{eq:hfr}
    \bar{h}(f) = \dfrac{8\pi^{2/3}}{\sqrt{10}}\dfrac{\left(G \mathcal{M}(1+z)\right)^{5/3}}{c^4 d_L}f^{2/3}. 
\ee
We stress that \cref{eq:hfr} enters only in the computation of the binned spectrum, which we use to validate the normalization of the population against \cref{eq:hcd2N}; the signal realizations entering our statistical analysis are instead built from the unaveraged amplitude $h(f)$ in the following way.
For a given expected number of sources, we first draw $\Ns$ from a Poisson distribution. The distribution $\d^3N/(\d z\d\mathcal{M} \d \ln f)$ can be factorized exactly into independent distributions for $\mathcal{M}$, $z$, and $\ln f$, which we sample by numerical inversion of the corresponding CDFs, restricting the frequency range to the bin under consideration.
Typically, we assume that the population is spatially isotropic. Therefore, for each binary, we draw the sky direction uniformly over the sphere; equivalently, $\phi \sim {\rm U}[0,2\pi)$ and $\cos\theta \sim {\rm U}[-1,1]$. Likewise, the orientation angles are sampled isotropically, with $\cos\iota \sim {\rm U}[-1,1]$ and polarization angle $\psi \sim {\rm U}[0,\pi)$. 
The polarization amplitudes are then built from \cref{eq:hplus,eq:hcross} using the unaveraged amplitude $h = \sqrt{5/2}\,\bar h$, with $\bar h$ given in \cref{eq:hfr}, and rotated by $\psi$ according to the transformation in \cref{eq:hpolplus,eq:hpolcross}. The resulting $\tilde h'_{P}$ are accumulated in the HEALPix pixel corresponding to the source direction, giving the pixel-space map that is subsequently contracted with the PTA response.
Our implementation \texttt{fastropop} can be found at \href{https://github.com/jonaselgammal/fastropop}{github.com/jonaselgammal/fastropop}.

\section{Kolmogorov-Smirnov test}\label{app:statistical_tests}

In this work, the statistics of the data are tested using the Kolmogorov-Smirnov (KS) test. In the following, we briefly review its definition and discuss subtleties related to its application to data whose variance is unknown.

Given a sample of $N$ data points $\vec{x} = \{x_1, \ldots, x_N\}$, the KS statistic measures the maximum discrepancy between the empirical CDF $F_N(x)$ and the hypothesized CDF $F(x)$:
\begin{equation}
D_{\mathrm{KS}} = \sup_{x}\,\bigl|F_N(x) - F(x)\bigr|.
\end{equation}
For a fully specified null (in our case, this would correspond to a known scale $\sigma^2$ of the underlying Gaussian distribution), the $p$-value follows the classical Kolmogorov--Smirnov distribution~\cite{Kolmogorov:1933,Smirnov:1948}. However, this distribution cannot be used directly when the scale is unknown and has to be estimated from the data as $\hat\sigma^2 = ( \sum_i |x_i|^2 ) / (2 N ) $. In this case, the standard $Q_{\mathrm{KS}}$ formula is \emph{anti-conservative}: the null distribution of $D_{\mathrm{KS}}$ is systematically smaller because the fitted distribution is pulled toward the data. This is known as the Lilliefors problem~\cite{Lilliefors:1967,Stephens:1974}. 

To define an adjusted KS test that correctly accounts for this effect, one can perform a bootstrap calibration~\cite{Efron:1993}. In practice, we proceed as follows:  
\begin{enumerate}
  \item For any given sample $\vec{x}$, estimate $\hat\sigma^2$ using $\hat\sigma^2 = ( \sum_i |x_i|^2 ) / (2 N )$, define $\vec{y} \equiv \vec{x} / \hat\sigma$ and compute $D_{\mathrm{KS}}$ taking $F(x)$ to be the CDF for the unit-scale null (i.e., normalizing out the scale). 
  \item Generate bootstrap samples $\vec{x}_i$ with $i \in \{ 1, \dots, N_{\rm boot} \} $ with unit scale.
  \item For each bootstrap sample, estimate the scale $\hat\sigma^{2}_i$, compute $\vec{y}_i = \vec{x}_i  / \hat\sigma_i$ and compute $D_{\mathrm{KS}, i}$ as in point 1.
  \item The $p$-value is the fraction of bootstrap statistics exceeding the observed $D_{\mathrm{KS}}$.
\end{enumerate}
In practice, we set $N_{\rm boot} = 2 \times 10^4$. 

\section{Single source case}\label{app:single_source}
For a single SMBHB, the maximally NG limit, the independence of the PDFs from the source distribution can be derived analytically. In a single frequency bin dominated by one source, the whitened data reads 
\begin{equation}
  w_I = c_I h_0 e^{i\varphi_0},
  \qquad
  |w_I|^2 = |c_I|^2 h_0^2,
\end{equation}
where $c_I$ depends only on the PTA geometry and source sky position, $h_0$ is the real-valued amplitude, and $\varphi_0$ an overall phase. Estimating the scale from the same sample,
\begin{equation}
  \hat{\sigma}^2 = \frac{1}{\Np}\sum_{I=1}^{\Np} |w_I|^2,
\end{equation}
the normalized data become
\begin{equation}
  \frac{|w_I|^2}{\hat{\sigma}^2}
  =
  \frac{|c_I|^2}{\frac{1}{\Np}\sum_J |c_J|^2},
\end{equation}
and the random factor $|h_0|^2$ cancels exactly. The test thus probes not the distribution of $h_0$, but only the PTA response for the specific realization. For $\Ns>1$ the cancellation is not exact, and the amplitude distributions remain slightly different.

\end{document}